\documentclass[final,5p,times,english,twocolumn]{elsarticle}

\usepackage{amssymb}
\usepackage{tensor}

\usepackage{amsmath}
\usepackage{xcolor}
\usepackage{subcaption}
\usepackage{dsfont}
\usepackage{mhchem}
\usepackage{background}

\journal{arXiv}

\usepackage{hyperref}
\hypersetup{
     colorlinks=true,
      linkcolor=blue,
}

\DeclareMathOperator{\arsinh}{arsinh}
\DeclareMathOperator{\erf}{erf}

\backgroundsetup{contents=}
\begin{document}

\begin{frontmatter}



\title{Reply to "Comments to Marvel Fusions Mixed Fuels Reactor Concept"}


\author{Hartmut Ruhl and Georg Korn} 

\address{Marvel Fusion, Theresienh\"ohe 12, 80339 Munich, Germany}

\begin{abstract}
In arXiv:2312.13429 Lackner et al. use standard methods to decide if it is possible to ignite mixed fuels. They correctly identify that the increased radiation losses make ignition significantly more challenging than for pure DT fuels, since this leads to higher ignition temperatures. Further, they conclude that at those temperatures the reduced electronic $\alpha$-stopping makes ignition impossible. We show that this conclusion is not correct. The model used for $\alpha$-stopping by Lackner et al. is only approximately correct for low temperatures and hydrogen isotopes. By extending the $\alpha$-stopping model to include ionic $\alpha$-stopping we show in \cite{ruhlkornarXiv5} that the contribution of ionic $\alpha$-particle stopping cannot be neglected. The ionic $\alpha$-stopping together with the neutron stopping, which is also neglected by Lackner et al., lead to elevated ion temperatures implying $kT_i > kT_e$. Those three effects combined lead us to the conclusion, that ignition of mixed fuels is indeed possible with far reaching implications, contrary to the analysis by Lackner.
\end{abstract}


\end{frontmatter}


\tableofcontents

\section{Introduction}
In a series of publications \cite{ruhlkornarXiv0, ruhlkornarXiv1, ruhlkornarXiv2, ruhlkornarXiv3,ruhl2025properties}, we explore the use of mixed fuels for inertial fusion applications. Our work has garnered interest from Lackner et al. \cite{lackner2023comments}, who investigated whether mixed fuels can be ignited at all. While mixed fuels have been studied far less than pure DT concepts, the question of their ignition has previously been addressed by Guskov et al. \cite{gus2013compression}, which serves as the foundation for our own analysis \cite{ruhlkornarXiv5}. Our conclusions align qualitatively with those of Guskov et al. \cite{gus2013compression} finding that mixed fuels can ignite but diverge from the findings of Lackner et al. \cite{lackner2023comments} claiming that mixed fuels cannot ignite. This discrepancy arises because we extend the analysis of Guskov et al. by incorporating additional physical effects that were overlooked in the work of Lackner et al. The analysis presented by Lackner et al. relies on textbook formulas from Atzeni and Meyer-ter-Vehn \cite{atzeni2004physics}. While analyses based on standard textbooks are generally reliable, it is crucial to ensure that all underlying assumptions of these formulas are applicable in the new context. We believe this is not the case for the stopping power method employed by Lackner et al. In the following section, we will detail the specific assumptions made in \cite{atzeni2004physics} that are not valid for mixed fuels.

\section{Ionic Stopping}
Some of the approximations in the $alpha$-stopping model employed by \cite{lackner2023comments} cannot be used for mixed fuels at the temperatures relevant. To show this we briefly sketch the derivation of the relevant stopping 
power model (see \cite{ruhlkornarXiv5}). The friction force between two species with charges \(q_b,q_c\), masses \(m_b,m_c\), number densities \(n_b,n_c\), temperatures \(T_b,T_c\) and relative velocity \(\bold{u}_{bc}\) and Coulomb logarithm \(\ln \Lambda_{bc}\) is given by
\begin{align}
\label{frictionforce}
\bold{R}_{bc}=-\frac{q_b^2 q_c^2 n_b n_c \ln \Lambda_{bc}}{(2\pi)^{3/2} \epsilon_0^2 \mu_{bc} } \, 
\frac{\bold{u}_{bc} }{u_{bc}^3}
 \left( \, \sqrt{\frac{\pi}{2}}   \mathrm{erf}\left(\frac{u_{bc}}{
\sqrt{2\left(\frac{kT_b}{m_b}+\frac{kT_c}{m_c}\right)}}\right)\right.\nonumber\\ 
\left.-\frac{u_{bc}}{
\sqrt{\left(\frac{kT_b}{m_b}+\frac{kT_c}{m_c}\right)}}\exp\left(-\frac{u_{bc}^2}{
2\left(\frac{kT_b}{m_b}+\frac{kT_c}{m_c}\right)}\right)\right).
\end{align}
For the case of fusion born alpha particles getting stopped in a hot plasma, there are contributions from each species to the total friction force, one from the electrons and one for each ion species. For low temperatures, the electron stopping dominates, since for the reduced mass in the denominator it holds $\mu_{e\alpha}\ll \mu_{i\alpha}$. For that reason the ionic contributions to the $\alpha$-stopping are neglected completely. Next, since $\frac{kT_e}{m_e}+\frac{kT_\alpha}{m_\alpha}\approx\frac{kT_e}{m_e}$ and also $u_{e\alpha}\ll \sqrt{\frac{2kT_e}{m_e}}$ one obtains
\begin{align}
    R_{e\alpha }\approx\frac{q_e^2 q_\alpha^2 n_e n_\alpha \ln \Lambda_{e\alpha}}{(2\pi)^{3/2} \epsilon_0^2 \mu_{e\alpha} } \, 
\frac{1}{3}\left(\sqrt{\frac{m_e}{kT_e}}\right)^3u_{e\alpha}\,.
\end{align}
This $T^{-3/2}$ scaling with the temperature, is what Lacker et al. call a particular negative synergy of the loss channels and it is central to their argument. Their analysis shows that the $\alpha$-transparency is too high at the increased ignition temperatures due to the increased radiation losses. The flaw in this argument is, that neglecting the ionic stopping contributions becomes increasingly wrong at higher temperatures. Lackner et al. say "At electron
temperatures $T_e\le 20keV$ more than 75\% of the $\alpha$- particle energy goes directly into heating the electrons [15]". What Lackner et al. fail to mention is that according to \cite{sigmar1971plasma}[their [15]] for $T_e\ge 20keV$ more than 25 \% of the $\alpha$- particle energy goes directly into the ions. This can be read off most conveniently in Fig. 3 of \cite{sigmar1971plasma}. Lackner et al. still use their flawed model for much higher temperatures than $20 keV$ and plot up to a temperature of $100 keV$. Even more importantly, \cite{sigmar1971plasma} as well as the analysis in \cite{atzeni2004physics} only consider hydrogen isotopes. But since the stopping power is proportional to $Z^2$ Boron ions for example have a 25 times increased stopping power. Additionally, for ions the other limiting case of \eqref{frictionforce} is relevant since $u_{i\alpha}\gg \sqrt{\frac{2kT_i}{m_i}+\frac{2kT_\alpha}{m_\alpha}}$ and in this limiting case the friction force is independent of the temperature. So, at increasing temperatures the relative contribution of the ionic stopping steadily increases. A detailed analysis of this effect can be found in \cite{ruhlkornarXiv5}. Lackner et al. only include the increased ion stopping power of mixed fuels due to the increased electron density, but the much more important $Z^2$ scaling of the ionic contribution is not taken into account. This $Z^2$ dependence of ionic stopping partially compensates the $n_e\sum_i n_i Z_i^2$ dependence for the radiation losses. Lackner et al. also do not consider neutron stopping, a second effect that directly heats the ions.

As a consequence of those two omissions, a third crucial effect is not part of Lackner's at al. analysis. Since ions do not lose energy directly but only by heat flow to electrons, any direct energy deposition into ions necessarily leads to $T_i>T_e$. This effect is very beneficial to fusion applications, since it leads to a much more favorable ratio between radiation losses and fusion energy gain. The analysis in \cite{ruhlkornarXiv5} shows that this effect makes the difference between ignition or no ignition at the confinement parameter $\rho R=0.5g/cm^2$. Lackner et al. consider the same confinement parameter.
Even without taking the ionic stopping or the temperatures gap between ions and electrons into account Guskov et al. \cite{gus2013compression} come to the conclusion that such fuels can ignite. We believe, that the analysis of Lackner et al. would also lead to such a conclusion, if they had taken higher $\rho R$ values into account. 

\section{Low Gain Consideration}
In this section, we explain our rationale for considering low-gain situations without ignition. It is evident that ignition is essential for a high-gain concept. However, experimentally achieving the conditions necessary for ignition is a significant challenge. Constructing a full-scale laser facility capable of providing these conditions requires substantial capital investment. Therefore, it is prudent to explore various aspects of the concept in experimentally feasible scenarios before committing to such a facility. In these experiments, we can only investigate situations with low 
$\rho R$ values and low gain. Even within these constraints, many aspects of the 
concept can be validated. This does not imply that we do not plan to eventually
achieve high $\rho R$ values and gains. This has seemingly led to some confusion. 
It is important to clarify that reaching a gain of $Q \approx 1$ does not require ignition. The timescale for the fuel to cool down allows for sufficient fusion 
reactions to achieve $Q \approx 1$. While this approach does not lead to a high-gain concept, such as hot spot ignition with a burn wave, it aligns with the objectives 
of our current investigations.

\section{Summary}
Standard analyses, as detailed in \cite{atzeni2004physics}, are often sufficient for many purposes. Even when they fall short, they still offer valuable guidance by clearly outlining the approximations made. The current issue serves as an instructive example. The discrepancy between \cite{ruhlkornarXiv5} and \cite{lackner2023comments} highlights the importance of periodically re-evaluating the approximations and simplifications that are commonly employed. In fact, mixed fuels represent a compelling research area that is more promising than it may initially appear.

\bibliographystyle{elsarticle-num} 
\bibliography{forensic1}

\end{document}